# Transmit Pulse Shaping for Molecular Communications

Siyi Wang, Weisi Guo[†], Mark D. McDonnell
Institute for Telecommunications Research, University of South Australia, Australia
[†]School of Engineering, The University of Warwick, United Kingdom
siyi.wang@mymail.unisa.edu.au, [†]weisi.guo@warwick.ac.uk, mark.mcdonnell@unisa.edu.au

*Abstract*—This paper presents a method for shaping the transmit pulse of a molecular signal such that the diffusion channel's response is a sharp pulse. The impulse response of a diffusion channel is typically characterised as having an infinitely long transient response. This can cause severe inter-symbol-interference, and reduce the achievable reliable bit rate. We achieve the desired chemical channel response by poisoning the channel with a secondary compound, such that it chemically cancels aspects of the primary information signal. We use two independent methods to show that the chemical concentration of the *information signal* should be $\propto \delta(t)$ and that of the *poison signal* should be $\propto t^{-3/2}$.

## I. INTRODUCTION

Molecular communications using nano-sized chemical particles is interesting to the field of biology and telecommunications. In terms of nature, there is significant interest in understanding the tether-less chemical communications channel between bacteria (microscopic scale) and between animals (macroscopic scale) [1]. In terms of engineering, there is also growing interest on these scales. For example, swarms of nano-robots can perform targeted drug delivery, and wireless sensors can be embedded in extremely challenging environments. In both of these cases, the communications challenge is that the local environment is often hostile to the utilization of radio-frequency (RF) systems. It has been shown in the last few years that chemical particles can be a viable carrier for information and they offer unique advantages and challenges [2].

In order to achieve general chemical-based communications, one needs to encode generic messages into a unique chemical pattern, and to reliably transport messages as continuous pulses. One of the challenges is the stochastic nature of the diffusion channel, which is a random walk process. The characteristic channel response an infinitely long transient response [3]–[6]. Therefore, the excess interference caused by a train of chemical pulses will build up at the receiver and cause significant bit errors. Being able to shape the transmit chemical pulse in such a way that the received pulse resembles a sharp pulse is vital and not well explored in literature. The difficulty in pulse shaping in chemistry is the challenging nature of the channel response, and the fact that chemical pulses must be strictly positive signals.

In this paper we assume the channel is a linear system and present two different methods of pulse shaping and show they achieve the same result, which is essentially an information bearing signal that is complemented by a poison signal to cancel the transient response.

## II. PULSE SHAPING FORMULATION

### A. Characterising the Channel

Let us consider a on-off-keying binary encoded system, where by a semi-infinite channel separates a chemical transmitter and a receiver. As shown in [2], [7], subject to an unit impulse input, the hitting concentration at distance $x$ is: $\phi_h(x,t) = \frac{1}{\sqrt{\pi Dt}} \exp\left(-\frac{x^2}{4Dt}\right)$. For a receiver that captures molecules and does not re-emit them, the number of molecules captured $\phi_c(x,t)$ is given by [2], [4]:

$$\phi_c(x,t) = \text{erfc}\left(\frac{x}{2\sqrt{Dt}}\right). \quad (1)$$

Eq. (1) is the cumulative function of the captured molecules, which monotonically increases with time $t$. To obtain the instantaneous number of molecules captured (impulse response $h(t)$), we differentiate Eq.(1) with respect to $t$:

$$h(t) = \frac{d\phi_c(x,t)}{dt}, \quad \mathscr{L}[h(t)] = H(s) = \exp\left(-\frac{x\sqrt{s}}{\sqrt{D}}\right). \quad (2)$$

The impulse response in Eq. (2) has been verified experimentally in our own experimentation [8]. In Fig. 1 (top), we show the impulse response with additive noise for a string of inputs. Clearly the combined effect of inter-symbol-interference (ISI) and additive noise is significant and the probability of incurring bit errors is high.

### B. *Method A:* Channel Inversion

We first consider inverting the channel in the complex s-domain. Let us consider a desired received pulse shape that is a sharp pulse $y_d(t) = \delta(t)$. Therefore, the desirable composite input signal is:

$$X_d(s) = \frac{Y_d(s)}{H(s)} = \exp\left(\frac{x\sqrt{s}}{\sqrt{D}}\right), \quad (3)$$

for $Y_d(s) = 1$.

If we consider a transmission system, where the distance is much smaller than the diffusivity rate ($x \ll \sqrt{D}$)[1], series expansion can be employed ($\exp(u) \approx 1 + u$). Inverse Laplace

---

[1]Typically found in either micro-biological environments (small $x$) or in high diffusivity channels (large $D$) [2].

transform yields two terms [2]:

$$x_{\rm d}(t) = \mathscr{L}^{-1}[X_{\rm d}(s)] \approx \delta(t) - \frac{x}{2\sqrt{D\pi}} t^{-3/2}. \quad (4)$$

The composite transmit pulse derived in Eq. (4) has a positive element ($t = 0$) and a negative element ($t > 0$). Let $\delta(t)$ be the information signal (compound A), and the $\propto t^{-3/2}$ term be the poison signal (compound B) which can cancel out aspects of compound A's impulse response.

### C. *Method B: Design Window Function*

Alternatively, we can consider windowing the impulse response $h(t)$ in the time-domain, such that we design the poison signal's response first. This is the reverse of finding the poison signal's pulse shape at the transmitter side in Method A. Sensible attributes for the poison signal's response include: i) zero value from $0 \leqslant t < T$; ii) the same tail shape as $h(t)$ from $t \geqslant T$. The value $T$ should be chosen such that it is greater than the time corresponding to the peak value of the impulse response ($T > t_{\max} = \frac{x^2}{6D}$). Given these parameters, the poison signal's transient response ($p_{\rm r}(t)$) can be defined as a *windowed* version of the impulse response:

$$p_{\rm r}(t) = h(t)u(t-T). \quad (5)$$

In order to find the desired poison pulse at the transmitter, we de-convolute $p_{\rm r}(t)$ with the channel:

$$\begin{aligned} P_{\rm d}(s) &= \frac{\mathscr{L}[p_{\rm r}(t)](s)}{H(s)}, \\ &= \frac{1}{2}\left[1 + {\rm erf}\left(1 - \frac{x\sqrt{s}}{2\sqrt{D}}\right) - e^{\frac{2x\sqrt{s}}{\sqrt{D}}} {\rm erfc}\left(1 + \frac{x\sqrt{s}}{2\sqrt{D}}\right)\right], \\ &\approx {\rm erf}(1) - \frac{x\sqrt{s}}{\sqrt{D}} {\rm erfc}(1), \end{aligned} \quad (6)$$

by using series expansion for $x \ll \sqrt{D}$ and set $T = \frac{3}{2} t_{\max}$.

Combining with the primary information signal ($x(t) = \delta(t)$), the composite signal is:

$$x_{\rm d}(t) = \delta(t) - \mathscr{L}^{-1}[P_{\rm d}(s)] \approx \Xi\left[\delta(t) - \frac{x}{\sqrt{D}} t^{-3/2}\right], \quad (7)$$

where $\Xi = {\rm erfc}(1)$.

### D. *Results and Discussion*

The composite transmit pulse derived in both Eq. (4) and Eq. (7) have a positive information element ($t = 0$) and a negative poison element that is continuously emitted ($t > 0$). We further note that the pulse shape derived using the windowing method in Eq. (7) is a scaled version of that derived via the channel inversion method in Eq. (4). In verifying our results, Fig. 1 shows the simulated system response with additive noise. Clearly the poisoned system response will yield fewer errors and be able to achieve a higher reliable data rate. The pulse

---

[2] The pseudo inverse Laplace transform of $\sqrt{s}$ is given in [9], but there is not a forward transform available due to the divergent nature at $t = 0$.

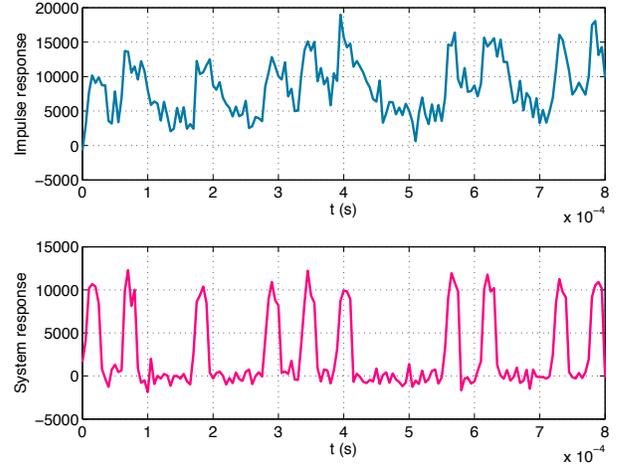

Fig. 1. System response (concentration) with additive noise for: (top) impulse input, and (bottom) impulse and poison input. The binary input is: $\{1,1,0,1,0,1,1,1,0,0,1,1,0,1,1\}$, $\frac{x}{\sqrt{D}} = 0.01$ and SNR = 15 dB

shapes derived are valid for all microscopic or high diffusivity chemical channels. In reality, how we emit the signal derived in this paper is not obvious, given its divergent and undefined nature at $t = 0$. This is a problem for future research to find practical pulse shapes.

## III. CONCLUSIONS

The paper has presented a simple and effective method of shaping a sharp pulse response to a chemical diffusion channel for communications. The difficulty was the challenging nature of the diffusion channel and that chemical pulses must be strictly positive signals. We employed two methods to show that the *information signal* should be $\propto \delta(t)$ and a further *poison signal* $\propto t^{-3/2}$ is required to chemically cancel the information signal.

## REFERENCES

[1] T. D. Wyatt, "Fifty years of pheromones," *Nature*, vol. 457, no. 7227, pp. 262–263, Jan. 2009.
[2] T. Nakano, A. Eckford, and T. Haraguchi, *Molecular Communication*. Cambridge University Press, 2013.
[3] K. Srinivas, A. Eckford, and R. Adve, "Molecular communication in fluid media: The additive inverse gaussian noise channel," *IEEE Transactions on Information Theory*, vol. 58, no. 7, pp. 4678–4692, Jul. 2012.
[4] W. Guo, S. Wang, A. Eckford, and J. Wu, "Reliable communication envelopes of molecular diffusion channels," *IET Electronics Letters*, vol. 49, no. 19, pp. 1248–1249, Sep. 2013.
[5] B. Atakan and O. B. Akan, "An information theoretical approach for molecular communication," in *IEEE Bio-Inspired Models of Network, Information and Computing Systems*, Dec. 2007, pp. 33–40.
[6] M. Pierobon and I. F. Akyildiz, "A physical end-to-end model for molecular communication in nanonetworks," *IEEE Journal on Selected Areas in Communications*, vol. 28, no. 4, pp. 602–611, 2010.
[7] S. M. Ross, *Stochastic Processes*. John Wiley, 1996.
[8] N. Farsad, W. Guo, and A. W. Eckford, "Tabletop molecular communication: text messages through chemical signals," *PLOS ONE*, vol. 8, no. 12, p. e82935, Dec. 2013.
[9] J. Abate and P. Valkó, "Multi-precision laplace transform inversion," *Wiley International Journal for Numerical Methods in Engineering*, vol. 60, no. 5, pp. 979–993, May 2004.